# DESCRIBING DISABILITY THROUGH INDIVIDUAL-LEVEL MIXTURE MODELS FOR MULTIVARIATE BINARY DATA[1]

By Elena A. Erosheva, Stephen E. Fienberg
and Cyrille Joutard

*University of Washington, Carnegie Mellon University
and Université Toulouse 1*

Data on functional disability are of widespread policy interest in the United States, especially with respect to planning for Medicare and Social Security for a growing population of elderly adults. We consider an extract of functional disability data from the National Long Term Care Survey (NLTCS) and attempt to develop disability profiles using variations of the Grade of Membership (GoM) model. We first describe GoM as an individual-level mixture model that allows individuals to have partial membership in several mixture components simultaneously. We then prove the equivalence between individual-level and population-level mixture models, and use this property to develop a Markov Chain Monte Carlo algorithm for Bayesian estimation of the model. We use our approach to analyze functional disability data from the NLTCS.

## 1. Introduction.

1.1. *Background.* Data on functional disability are of widespread policy interest in the United States, especially with respect to planning Medicare and Social Security spending for a growing population of elderly adults. The concept of functional disability reflects difficulties in performing activities that are considered normal for everyday living. These activities are usually divided into two types, namely, basic and instrumental activities of daily

Received February 2007; revised June 2007.
[1]Supported in part by NIH Grant R01 AG023141-01 to Carnegie Mellon University, and by the Center of Statistics and the Social Sciences, University of Washington, under a Seed Grant to Elena A. Erosheva.
Supplementary material available at http://imstat.org/aoas/supplements

living (ADL and IADL). ADL and IADL outcomes are considered essential in health services research and form a cornerstone of geriatric medicine. In this article we present a Bayesian analysis of functional disability among a sample of elderly individuals in the National Long Term Care Survey (NLTCS), using basic and extended Grade of Membership (GoM) models for multivariate binary response data.

The NLTCS began in 1982 and now extends over six waves through 2004, making it an important source of information on possible changes in disability over time among the elderly Americans. The NLTCS data on functional disability have been used to generate some major findings, such as a persistent decline in chronic disability among the elderly Americans [Manton and Gu (1999) and Manton, Gu and Lamb (2006a, 2006b)].

It is common practice to analyze functional disability data by using totals where individual scores are added together for all items or by subsets [Manton and Gu (1999)]. Statistically, adherence to the Rasch model (1960) can provide researchers with a formal justification for reducing the multivariate data down to such total scores. It is often the case, however, that functional disability data have a high amount of heterogeneity that is not explainable by the Rasch model. It may be possible to circumvent this problem by reducing the set of functional disability items under consideration, as was illustrated, for example, in the gerontology literature by Spector and Fleishman (1998). This approach, however, obviously ignores potentially relevant information contained in the excluded items.

In this paper we use individual-level mixtures to account for heterogeneity in functional disability data measured with a given battery of items without considering the issue of item reduction. We contrast the individual-level mixture assumption with population-level mixture models that assume individuals can be members of one and only one subpopulation, such as latent class models [Goodman (1974), Lazarsfeld and Henry (1968)]. The central idea of all individual-level mixture models is to allow an individual's membership to be a mixture with respect to population components [Blei, Jordan and Ng (2003), Pritchard, Stephens and Donnelly (2000), Woodbury, Clive and Garson (1978)]. A natural example of individual-level mixtures is genetic makeup of individuals who have various degrees of ancestry in several subpopulations of origin [Pritchard, Stephens and Donnelly (2000)]. Such admixed individuals do not simply belong to one of the original subpopulations with some degree of uncertainty, but their genetic makeup is actually composed of genes that originated from different subpopulations. Specifically, we use the Grade of Membership (GoM) model introduced in 1978 by Woodbury, Clive and Garson (1978) and develop its extension to address the following questions: How many mixture categories are in the functional disability data under the assumption of mixed membership? What are characteristics of each mixture category? What is the population distribution of the individual membership scores?

We begin by introducing the NLTCS in Section 2. Next, we describe the GoM model and its relationship to latent class models via the fundamental representation theorem in Section 3. We use this result to develop a fully Bayesian approach in Section 4.1, and describe a variational approximation approach as an alternative estimation method in Section 4.2. Section 5 develops the extended mixture GoM model and corresponding estimation techniques. Section 6 considers the question of dimensionality selection in terms of the optimal number of mixture categories. Section 7 describes results from simulation studies. Finally, we present an individual-level mixture analysis of the NLTCS functional disability data and provide discussions in Sections 8 and 9.

**2. National Long Term Care Survey functional disability data.** The NLTCS aims to assess chronic disability in the U.S. Medicare-enrolled population age 65 or older [Corder and Manton (1991)]. The survey began in 1982 with a screening survey instrument that selected community-dwelling chronically disabled (based on basic and instrumental activities of daily living) persons for detailed in-home interviews. Once individuals screened-in, the NLTCS followed them longitudinally. The second wave of the survey was in 1984, and all subsequent waves occurred in five-year intervals with the most recent wave completed in 2004. The NLTCS replenishes its sample at each wave in order to reflect the current U.S. population 65 and older. While additional components have come and gone from post-1982 waves of the NLTCS, key disability questions have stayed the same. For more information on the NLTCS, see Corder and Manton (1991), Manton, Corder and Stallard (1997), Singer and Manton (1998).

We consider an extract from the NLTCS that contains data on 6 activities of daily living (ADL) and 10 instrumental activities of daily living (IADL) for community-dwelling elderly from 1982, 1984, 1989 and 1994 survey waves. These 16 binary functional disability measures are described in detail in Manton, Corder and Stallard (1993). The 6 ADL items include basic activities of hygiene and personal care (eating, getting in/out of bed, getting around inside, dressing, bathing, and getting to the bathroom or using toilet). The 10 IADL items include basic activities necessary to reside in the community (doing heavy housework, doing light housework, doing laundry, cooking, grocery shopping, getting about outside, travelling, managing money, taking medicine and telephoning). Positive responses are coded as 1 = disabled, and negative as 0 = healthy. In the NLTCS, positive ADL responses mean that during the past week the activity had not been, or was not expected to be, performed without the aid of another person or the use of equipment; positive IADL responses mean that a person usually could not, or was not going to be able to, perform the activity because of a disability of a health problem. For a more in-depth discussion, see Manton, Corder and Stallard (1993) and Erosheva and White (2006).

At each wave, the survey sample is representative of the 65 years and older U.S. population at that point in time. High follow-up rates and consistency in ADL and IADL questions over time make the NLTCS a unique source of data for studying complex questions such as the dynamics of population changes in disability. Manton and Gu (1999) and Manton, Gu and Lamb (2006a, 2006b) used weighted total numbers of impaired ADL and IADL to show declines in disability, but the important question of "Why?" [Cutler (2001)] still remains open. We believe that in order to move forward in our understanding of why disability is declining so rapidly and whether the decline can be expected to continue, an important first step is to describe heterogeneous multivariate disability manifestations.

Our ultimate goal is to develop a longitudinal version of the GoM model. Our analysis in this paper represents an attempt to learn disability mixture profiles that describe the underlying structure of functional disability in the chronically disabled community-dwelling elderly U.S. population. We make three simplifying assumptions in this analysis. First, we assume that the nature of the mixture components stays the same over time. For a longitudinal version of the GoM model, keeping profiles the same over time and allowing the population distribution among the profiles to change would allow us to obtain an estimable model. For similar reasons, the assumption of time-invariant latent classes is common in latent class transition modeling [see Reboussin et al. (1998), e.g.]. In addition, our exploratory analysis where we analyzed each wave separately using the GoM model, yielded profiles whose characteristics were fairly stable over time, thus confirming that the assumption of profile time-invariance is reasonable in our case. Second, we assume no inter-dependencies between longitudinal records on the same individuals. Violations of this assumption may affect efficiency of our estimates but will not introduce bias. Third, we ignore the sample weights associated with differential probabilities of selection into the NLTCS. In fact, we have yet to understand how if at all we could incorporate the weights into the modeling process. We view these three assumptions necessary for this first step toward understanding changes in disability over time.

**3. The grade of membership (GoM) model and its latent class representation.** The GoM model originate in the context of medical applications: when a diagnosis is uncertain, partial membership reflects this uncertainty through allowing different disease symptoms to correspond to different stages of the disease. GoM applications now cover a wide spectrum of studies, ranging from studying depression [Davison et al. (1989)] and schizophrenia [Manton et al. (1994)] to analyzing complex genotype-phenotype relations [Manton et al. (2004)]; for a recent review, see Erosheva and Fienberg (2005). The model remains relatively unfamiliar to statistical audiences, however. Despite a multitude of published large-scale GoM applications, there are few statistical publications that explore basic GoM properties and demonstrate the model's utility with similar examples [Erosheva (2005), Potthoff (2000), Wachter (1999)].

In particular, the relationship between individual-level and population-level mixture models does not appear to be clearly formulated in the literature. Singer (1989) describes the GoM model as a new type of model that is not equivalent to usual mixture models. Likewise, when comparing the GoM and latent class models, in their 1994 book, Manton, Woodbury and Tolley (1994) concluded: "latent class model is nested in the GoM model structure...," but "...if we allow latent class model to have more classes, then it is potentially possible to "fit" the realized data set as well as with GoM" (page 45). On the other hand, Haberman (1995) in his review of Manton et al., suggested that the GoM model is a special case of latent class models. He pointed out that a set of constraints imposed upon a latent class model can specify a distribution of manifest variables that is identical to that specified by the GoM model.

In this Section we describe the GoM and latent class models and present the fundamental representation theorem of equivalence between individual-level and population-level mixture models [Erosheva (2006)].

*GoM and latent class models.* Let $x = (x_1, x_2, \ldots, x_J)$ be a vector of polytomous manifest variables, where $x_j$ takes on values $l_j \in \mathcal{L}_j = \{1, 2, \ldots, L_j\}$, $j = 1, 2, \ldots, J$, and $L_j$ denotes the number of possible outcomes. Let $\mathcal{X} = \prod_{j=1}^{J} \mathcal{L}_j$ be the set of all possible outcomes for vector $x$.

To define the GoM model, let $K$ be the number of mixture components (extreme profiles), and let $g = (g_1, g_2, \ldots, g_K)$ be a latent partial membership vector of $K$ nonnegative random variables that sum to 1. For discrete data, each extreme profile is characterized by a vector of conditional response probabilities, when a given $k$th component of the partial membership vector is 1 and the others are 0:

(1)
$$\lambda_{kjl_j} = \mathrm{pr}(x_j = l_j | g_k = 1), \qquad \begin{aligned} &k = 1, 2, \ldots, K, \\ &j = 1, 2, \ldots, J, \\ &l_j = 1, 2, \ldots, L_j. \end{aligned}$$

The set of conditional response probabilities must satisfy the following constraints:

$$\sum_{l_j \in \mathcal{L}_j} \lambda_{kjl_j} = 1, \qquad k = 1, 2, \ldots, K; j = 1, 2, \ldots, J.$$

Given partial membership vector $g \in [0, 1]^K$, the conditional distribution of manifest variable $x_j$ is given by a convex combination of the extreme profiles' conditional response probabilities, that is,

$$(2) \qquad \mathrm{pr}(x_j = l_j | g) = \sum_{k=1}^{K} g_k \lambda_{kjl_j}, \qquad j = 1, 2, \ldots, J, l_j = 1, 2, \ldots, L_j.$$

The local independence assumption states that manifest variables are conditionally independent, given latent variables. Under this assumption, the conditional probability of observing response pattern $l$ is

$$f^{\mathrm{GoM}}(l | g) = \mathrm{pr}(x = l | g)$$

$$= \prod_{j=1}^{J} \mathrm{pr}(x_j = l_j | g) = \prod_{j=1}^{J} \left( \sum_{k=1}^{K} g_k \lambda_{kjl_j} \right), \qquad l \in \mathcal{X}.$$

The local independence assumption is common in latent structure models [Lazarsfeld and Henry (1968)]; it says that latent variables fully account for associations among the observed responses.

Let us denote the distribution of $g$ by $D(g)$. Integrating out the latent variable $g$, we obtain the marginal distribution for response pattern $l$ in the form of an individual-level mixture

$$f^{\mathrm{GoM}}(l) = \mathrm{pr}(x = l) = \int f^{\mathrm{GoM}}(l | g) \, dD(g)$$

$$(3)$$

$$= \int \prod_{j=1}^{J} \left( \sum_{k=1}^{K} g_k \lambda_{kjl_j} \right) dD(g), \qquad l \in \mathcal{X}.$$

Using similar notation, we can derive the $K$-class population-level mixture (latent class) model as a special case of the $K$-profile GoM model by restricting components of the partial membership vector to only take values 0 and 1. Denote the restricted version of the membership vector by $g^*$ and its probability mass function by $\pi_k = \mathrm{pr}(g_k^* = 1)$. Assuming local independence, we see that the marginal distribution of the manifest variables under the latent class model simplifies to the $K$-component summation:

$$f^{\mathrm{LCM}}(l) = \mathrm{pr}(x = l) = \int f^{\mathrm{LCM}}(l | g^*) \, dD(g^*)$$

$$(4)$$

$$= \sum_{k=1}^{K} \pi_k \prod_{j=1}^{J} \lambda_{kjl_j}, \qquad l \in \mathcal{X}.$$

The probability of observing response pattern $l$ is the sum of the probabilities of observing $l$ from each of the latent classes, weighted by their relative sizes, $\pi_k$. One can visualize the relationship between sets of individual-specific response probabilities under the GoM and latent class models with the same number of mixture categories using a geometric approach [Erosheva (2005)].

*Fundamental representation theorem.* Note that the GoM marginal or integrated likelihood in equation (3) does not simplify to a summation of $K$ components. This is in contrast to the functional form of the likelihood for a population-level mixture of $K$ latent classes in equation (4). If we relax the requirement of equality of the number of latent classes and extreme profiles, however, following Haberman (1995), we can construct a latent class model such that its marginal distribution of manifest variables is exactly the same as that under the GoM model.

Consider a vector of $J$ polytomous latent variables $z = (z_1, z_2, \ldots, z_J)$, each taking on values from the set of integers $\{1, 2, \ldots, K\}$. Vector $z$ here is the latent classification variable. Denote by $\mathcal{Z} = \{1, 2, \ldots, K\}^J$ the set of all possible vectors $z$. As before, $\mathcal{X} = \prod_{j=1}^{J} \mathcal{L}_j$ is the set of all possible outcomes for vector $x$. Then $\mathcal{X} \times \mathcal{Z}$ is the index set for the cross-classification of the manifest variables $x$ and latent classification variables $z$.

To obtain a latent class representation of the GoM model, we must find a way to interchange the summation and the product operator in equation (3). The following lemma provides algebra which allows us to do so.

LEMMA 3.1. *For any two positive integers $J$ and $K$, and for any two sets of real numbers $\{a_k, k = 1, 2, \ldots, K\}$ and $\{b_{kj}, k = 1, 2, \ldots, K, j = 1, 2, \ldots, J\}$,*

$$\prod_{j=1}^{J} \sum_{k=1}^{K} a_k b_{kj} = \sum_{z \in \mathcal{Z}} \prod_{j=1}^{J} a_{z_j} b_{z_j j},$$

*where $z = (z_1, z_2, \ldots, z_J)$ is such that $z \in \mathcal{Z} = \prod_{j=1}^{J} \{1, 2, \ldots, K\}$.*

Define the distribution over latent classes $z \in \mathcal{Z}$, conditional on the distribution of membership vector $g \in [0, 1]^K$:

$$(5) \qquad \pi_z = E_D \left( \prod_{j=1}^{J} g_{z_j} \right).$$

If $(g_1, g_2, \ldots, g_K)$ has a joint distribution $D(g)$ on $[0, 1]^K$, such that $g_1 + g_2 + \ldots + g_K = 1$, then $\pi_z$ is a probability measure on $\mathcal{Z}$. From the functional form of $\pi_z$, it also follows that latent classification variables $z_1, z_2, \ldots, z_J$ are exchangeable.

To specify the conditional distribution of the manifest variables given the latent variables $z$, we need two additional assumptions. First, assume that $x_j$ depends only on the $j$th component of the latent indicator variable $z$:

$$(6) \qquad \mathrm{pr}(x_j = l_j | z) = \mathrm{pr}(x_j = l_j | z_1, z_2, \ldots, z_J) = \mathrm{pr}(x_j = l_j | z_j),$$

where $z_j \in \{1, 2, \ldots, K\}$, and $l_j \in \{1, \ldots, L_j\}$ is the observed value of manifest variable $x_j$. In essence, equation (6) postulates that manifest variable $x_j$ is directly influenced only by the $j$th component of the latent classification vector $z$. Second, assume that conditional response probabilities in equation (6) are given by

$$
(7) \qquad \mathrm{pr}(x_j = l_j | z_j) = \lambda_{z_j j l_j}, \qquad
\begin{aligned}
& z_j = 1, 2, \ldots, K, \\
& j = 1, 2, \ldots, J, \\
& l_j = 1, 2, \ldots, L_j,
\end{aligned}
$$

where the set of $\lambda$s is the same as the set of conditional response probabilities for the GoM model. These structural parameters must also satisfy the constraints:

$$
\sum_{l_j=1}^{L_j} \lambda_{z_j j l_j} = 1, \qquad \text{for all } z \in \mathcal{Z}, j \in \{1, 2, \ldots, J\}.
$$

Under the local independence assumption, we obtain the probability of observing response pattern $l$ for the latent class model as

$$
(8) \qquad f^*(l) = \sum_{z \in \mathcal{Z}} \pi_z \left( \prod_{j=1}^{J} \lambda_{z_j j l_j} \right), \qquad l \in \mathcal{X},
$$

where the probability of latent class $z$ is the expected value of a $J$-fold product of the membership scores $\pi_z = E_D(\prod_{j=1}^{J} g_{z_j})$. Thus, the probability of observing response pattern $l$ in equation (8) is the sum of the conditional probabilities of observing $l$ from each of the latent classes, weighted by the latent class probabilities.

Consider the marginal probability of an arbitrary response pattern $l \in \mathcal{X}$ for the GoM model provided by equation (3). Applying lemma 3.1 with $a_k = g_k$, $b_{kj} = \lambda_{kjl_j}$, and using properties of expectation, we obtain the marginal probability:

$$
f^{\mathrm{GoM}}(l) = \sum_{z \in \mathcal{Z}} \left\{ E_D \left( \prod_{j=1}^{J} g_{z_j} \right) \left( \prod_{j=1}^{J} \lambda_{z_j j l_j} \right) \right\},
$$

which is exactly the same as in equation (8). It follows that the GoM model is equivalent to a latent class model with a distribution on the latent classes given by a functional form of the distribution of membership scores. This equivalence statement can be generalized via the following *fundamental representation theorem*:

THEOREM 3.2. *Given $J$ manifest variables, any individual-level mixture model with $K$ components can be represented as a constrained population-level mixture model with $K^J$ components.*

The fundamental representation theorem applies to a wider class of mixed membership models introduced by Erosheva (2002).

## 4. Estimation algorithms for the standard GoM model.

### 4.1. *Bayesian estimation algorithm.*

*Data augmentation.* The fundamental representation theorem leads us naturally to a data augmentation approach in the spirit of those described by Tanner ([1996](#)). In this Section we present the Bayesian estimation algorithm for the GoM model, described earlier in Erosheva ([2003](#)).

Denote by $\mathbf{x}$ the set of observed responses $x_{ij}$ for all subjects. Denote by $\boldsymbol{\lambda}$ the set of conditional response probabilities. For the functional disability data, $\lambda_{kj} = \text{pr}(x_j = 1 | g_k = 1)$ is the probability of being disabled on activity $j$ for a complete member of extreme profile $k$. For subject $i$, augment observed responses with realizations of the latent classification variables $z_i = (z_{i1}, \ldots, z_{iJ})$. Denote by $\mathbf{z}$ the set of latent classifications $z_{ij}$ on all items for all individuals. In the following, we use notation $p(\cdot)$ to refer to both probability density and probability mass functions.

We assume the distribution of membership scores is Dirichlet with parameters $\alpha$. The joint probability model for the parameters and augmented data is

$$p(\mathbf{x}, \mathbf{z}, \mathbf{g}, \boldsymbol{\lambda}, \alpha) = p(\boldsymbol{\lambda}, \alpha) \cdot p(\mathbf{x}, \mathbf{z}, \mathbf{g} | \boldsymbol{\lambda}, \alpha)$$

$$= p(\boldsymbol{\lambda}, \alpha) \prod_{i=1}^{N} [p(z_i | g_i) p(x_i | \boldsymbol{\lambda}, z_i) \cdot D(g_i | \alpha)],$$

where

$$p(z_i | g_i) = \prod_{j=1}^{J} \prod_{k=1}^{K} g_{ik}^{z_{ijk}},$$

$$p(x_i | \boldsymbol{\lambda}, z_i) = \prod_{j=1}^{J} \prod_{k=1}^{K} (\lambda_{kj}^{x_{ij}} (1 - \lambda_{kj})^{1 - x_{ij}})^{z_{ijk}},$$

$$Dir(g_i | \alpha) = \frac{\Gamma(\sum_k \alpha_k)}{\Gamma(\alpha_1) \cdots \Gamma(\alpha_K)} g_{i1}^{\alpha_1 - 1} \cdots g_{iK}^{\alpha_K - 1},$$

and latent classification indicators $z_{ijk}$ are such that $z_{ijk} = 1$, if $z_{ij} = k$, and $z_{ijk} = 0$ otherwise.

We assume the prior on extreme profile response probabilities $\boldsymbol{\lambda}$ is independent of the prior on the hyperparameters $\alpha$. We further assume that the prior distribution of extreme profile response probabilities treats items and extreme profiles as independent. Thus,

$$(9) \qquad p(\boldsymbol{\lambda}, \alpha) = p(\alpha) \prod_{k=1}^{K} \prod_{j=1}^{J} p(\lambda_{kj}).$$

We take $p(\lambda_{kj})$ to be $Beta(\eta_1, \eta_2)$, and, for simplicity, in what follows we use $\eta_1 = \eta_2 = 1$.

If the hyperparameters $\alpha$ are known, it is possible to obtain complete conditional distributions and use standard software such as BUGS[1] to obtain a posterior distribution of the model parameters [Erosheva (2002)]. In reality, the hyperparameters are unlikely to be known and need to be estimated. Setting hyperparameters to some fixed values without prior knowledge may bias conclusions and affect model choice in individual-level mixture models [see the discussion in Airoldi et al. (2007)].

If we assume that the Dirichlet parameter vector $\alpha$ is unknown, we obtain samples from its posterior distribution via a Metropolis–Hastings step within the Gibbs sampler. Consider a reparameterization of $\alpha = (\alpha_1, \ldots, \alpha_K)$ with $\alpha_0 = \sum_{k=1}^{K} \alpha_k$ and $\xi = (\xi_1, \ldots, \xi_K)$, where $\xi_k = \alpha_k/\alpha_0$. Then components of vector $\xi$ reflect proportions of the item responses that belong to each mixture category, and $\alpha_0$ reflects the spread of the membership distribution. The closer $\alpha_0$ is to zero, the more probability is concentrated near the mixture categories; similarly, the larger $\alpha_0$ is, the more probability is concentrated near the population average membership score.

We assume that $\alpha_0$ and $\xi$ are independent since they govern two unrelated qualities of the distribution of the GoM scores. We also assume that the prior distribution on the GoM scores is independent of the prior distribution on the structural parameters. The joint distribution of the parameters and augmented data is

$$(10) \quad p(\boldsymbol{\lambda})p(\alpha_0)p(\xi)\left(\prod_{i=1}^{N} D(g_i|\alpha)\right)\prod_{i=1}^{N}\prod_{j=1}^{J}\prod_{k=1}^{K}(g_{ik}\lambda_{kj}^{x_{ij}}(1-\lambda_{kj})^{1-x_{ij}})^{z_{ijk}}.$$

In the absence of a strong prior opinion about hyperparameters $\alpha_0$ and $\xi$, we take the prior distribution $p(\xi)$ to be uniform on the simplex and $p(\alpha_0)$ to be a proper diffuse gamma distribution.

*Sampling from the posterior distribution.*

- Imputation step: We use a multinomial complete conditional distribution to obtain the $(m+1)$st draw of latent class indicator variables $z_{ij}$ for each $i = 1, \ldots, N$, $j = 1, \ldots, J$:

$$(11) \quad z_{ij}^{(m+1)} \sim Mult(1, p_1, \ldots, p_K), \qquad p_k \propto g_{ik}\lambda_{kj}^{x_{ij}}(1-\lambda_{kj})^{1-x_{ij}}.$$

- Posterior step:

---

[1]The Bayesian inference using Gibbs Sampling project software [Spiegelhalter et al. (1996)]. For details see **http://www.mrc-bsu.cam.ac.uk/bugs/**.

– Sampling $\lambda$. We use the complete conditional distribution to obtain the $(m+1)$st draw of conditional response probabilities $\lambda_{kj}, k = 1, \ldots, K, j = 1, \ldots, J$:

$$(12) \qquad \lambda_{kj}^{(m+1)} \sim Beta\left(1 + \sum_{i=1}^{N} x_{ij} z_{ijk}, 1 + \sum_{i=1}^{N} (z_{ijk} - x_{ij} z_{ijk})\right).$$

– Sampling $g$. We use the complete conditional distribution to obtain the $(m+1)$st draw of membership scores $g_i, i = 1, \ldots, N$:

$$(13) \qquad g_i^{(m+1)} \sim D\left(\alpha_1 + \sum_{j=1}^{J} z_{ij1}, \ldots, \alpha_K + \sum_{j=1}^{J} z_{ijK}\right).$$

– Sampling $\alpha_0$ and $\xi$. Here we require Metropolis–Hastings steps.

*Sampling $\alpha_0$.* Let the prior $p(\alpha_0)$ be Gamma$(\tau, \beta)$ with shape parameter $\tau$ and inverse scale parameter $\beta$. The full conditional distribution for $\alpha_0$, up to a constant of proportionality, is

$$
\begin{aligned}
(14) \qquad p(\alpha_0 | \cdots) \propto{}& \alpha_0^{\tau-1} \exp\left[-\left(\beta - \sum_{k=1}^{K} \xi_k \sum_{i=1}^{N} \log g_{ik}\right)\alpha_0\right] \\
&\times \left[\frac{\Gamma(\alpha_0)}{\Gamma(\xi_1 \alpha_0) \cdots \Gamma(\xi_K \alpha_0)}\right]^N,
\end{aligned}
$$

where $\cdots$ in $p(\alpha_0 | \cdots)$ stands for all other variables.

In order to obtain the $(m+1)$st draw of $\alpha_0$ with the Metropolis–Hastings algorithm, we

1. Draw a candidate point $\alpha_0^*$ from a proposal distribution $p(\alpha_0^* | \alpha_0^{(m)})$;
2. Calculate the proposal ratio

$$r_{\alpha_0} = \frac{p(\alpha_0^* | \cdots) p(\alpha_0^{(m)} | \alpha_0^*)}{p(\alpha_0^{(m)} | \cdots) p(\alpha_0^* | \alpha_0^{(m)})};$$

3. Assign $\alpha_0^{(m+1)} = \alpha_0^*$ with probability $\min\{1, r_{\alpha_0}\}$, otherwise assign $\alpha_0^{(m+1)} = \alpha_0^{(m)}$.

We take the proposal distribution $p(\alpha_0^* | \alpha_0^{(m)})$ to be gamma with the expected value set at the value of the last draw, $\alpha_0^{(m)}$, and the shape parameter $\omega > 1$. The inverse scale parameter for the proposal distribution is then $\omega / \alpha_0^{(m)}$, where $\omega$ plays the role of the tuning parameter for the Metropolis–Hastings step. The proposal ratio for the $(m+1)$st draw of $\alpha_0$ is the product of the likelihood component and the component that accounts for the asymmetric proposal distribution:

$$r_{\alpha_0} = r_{\alpha_0}^L \cdot r_{\alpha_0}^A,$$

where

$$r_{\alpha_0}^L = \left(\frac{\alpha_0^*}{\alpha_0^{(m)}}\right)^{\tau-1} \exp\left[-\left(\beta - \sum_{k=1}^K \xi_k \sum_{i=1}^N \log g_{ik}\right)(\alpha_0^* - \alpha_0^{(m)})\right]$$

$$\times \left[\frac{\Gamma(\alpha_0^*)\Gamma(\xi_1\alpha_0^{(m)})\cdots\Gamma(\xi_K\alpha_0^{(m)})}{\Gamma(\alpha_0^{(m)})\Gamma(\xi_1\alpha_0^*)\cdots\Gamma(\xi_K\alpha_0^*)}\right]^N,$$

$$r_{\alpha_0}^A = \left(\frac{\alpha_0^{(m)}}{\alpha_0^*}\right)^{2\omega-1} \exp[-\omega(\alpha_0^{(m)}/\alpha_0^* - \alpha_0^*/\alpha_0^{(m)})].$$

*Sampling $\xi$.* The full conditional distribution for $\xi$, up to a constant of proportionality, is

$$(15) \qquad p(\xi|\cdots) \propto \exp\left[\alpha_0 \sum_{k=1}^K \xi_k \sum_{i=1}^N \log g_{ik}\right]\left[\frac{\Gamma(\alpha_0)}{\Gamma(\xi_1\alpha_0)\cdots\Gamma(\xi_K\alpha_0)}\right]^N,$$

where $\cdots$ in $p(\xi|\cdots)$ stands for all other variables.

The Metropolis–Hastings sampling algorithm to obtain the $(m+1)$st draw of $\xi$ has three steps:

1. Draw a candidate point $\xi^*$ from a proposal distribution $p(\xi^*|\xi^{(m)})$;
2. Calculate the proposal ratio

$$r_\xi = \frac{p(\xi^*|\cdots)p(\xi^{(m)}|\xi^*)}{p(\xi^{(m)}|\cdots)p(\xi^*|\xi^{(m)})};$$

3. Assign $\xi^{(m+1)} = \xi^*$ with probability $min\{1, r_\xi\}$, otherwise assign $\xi^{(m+1)} = \xi^{(m)}$.

We chose the proposal distribution for $\xi$ to be $Dir(\xi^*|\eta K\xi_1^{(m)}, \ldots, \eta K\xi_K^{(m)})$. The proposal distribution is centered at the previous draw and has reasonably small variance for each component, $\xi_k^{(m)}(1-\xi_k^{(m)})/(\eta K+1)$. The proposal ratio for $\xi$ is

$$r_\xi = \exp\left[\alpha_0 \sum_{k=1}^K \sum_{i=1}^N \log g_{ik}(\xi_k^* - \xi_k^{(m)})\right]\left[\frac{\Gamma(\xi_1^{(m)}\alpha_0)\cdots\Gamma(\xi_K^{(m)}\alpha_0)}{\Gamma(\xi_1^*\alpha_0)\cdots\Gamma(\xi_K^*\alpha_0)}\right]^N$$

$$\times \frac{\Gamma(\eta K\xi_1^{(m)})\cdots\Gamma(\eta K\xi_K^{(m)})}{\Gamma(\eta K\xi_1^*)\cdots\Gamma(\eta K\xi_K^*)} \cdot \frac{(\xi_1^{(m)})^{\xi^*-1}\cdots(\xi_K^{(m)})^{\xi^*-1}}{(\xi_1^*)^{\xi^{(m)}-1}\cdots(\xi_K^*)^{\xi^{(m)}-1}},$$

where $\eta$ is a tuning parameter.

4.2. *Variational approximation.* Variational approximation methods provide an alternative estimation approach by approximating a joint posterior distribution when the likelihood is intractable [see Jordan et al. (1999)].

They assume the model parameters are unknown but fixed. For the GoM model, the integrated likelihood for an individual

$$(16) \qquad p(x|\alpha, \boldsymbol{\lambda}) = \int \prod_{j=1}^{J} \left( \sum_{k=1}^{K} g_k \lambda_{kj}^{x_j} (1 - \lambda_{kj})^{1-x_j} \right) D_\alpha(dg),$$

does not have a closed form solution (the individual index $i$ is omitted to simplify the notation). To compute the joint posterior distribution $p(g, z|x, \alpha, \boldsymbol{\lambda})$ of the GoM scores $g = (g_1, \ldots, g_K)$ and the latent classifications variables $z = (z_1, \ldots, z_J)$, we consider $N$ independent fully factorized joint distributions, one for each individual:

$$q(g, z|\gamma, \boldsymbol{\phi}) = q(g|\gamma) \prod_{j=1}^{J} q(z_j|\phi_j).$$

Here, $(\gamma, \boldsymbol{\phi})$ is a set of free variational parameters, where $\gamma = (\gamma_1, \ldots, \gamma_K)$ and $\boldsymbol{\phi}$ is the matrix $\phi_{jk}$, $j = 1, \ldots, J$, $k = 1, \ldots, K$. Assuming $q(g|\gamma) = Dir(g|\gamma)$ and $q(z_j|\phi_j) = Mult(1, \phi_{j1}, \ldots, \phi_{jK})$, we employ Jensen's inequality to approximate the log-likelihood by a lower bound which becomes a function of the variational parameters, $(\gamma, \boldsymbol{\phi})$.

We derive (pseudo) maximum likelihood estimates of the model parameters $(\alpha, \boldsymbol{\lambda})$ by using an approximate EM algorithm. In the E-step, we obtain values of variational parameters $(\gamma, \boldsymbol{\phi})$ that yield the tightest possible lower bound. In the M-step, we maximize the lower bound with respect to the parameters of the model, $(\alpha, \boldsymbol{\lambda})$.

Given the current estimates of the model parameters $(\alpha, \boldsymbol{\lambda})$, the E step consists of updates:

$$(17) \qquad \phi_{jk} \propto \lambda_{kj}^{x_j} (1 - \lambda_{kj})^{1-x_j} \times \left( \Psi(\gamma_k) - \Psi\left( \sum_{k=1}^{K} \gamma_k \right) \right),$$

$$(18) \qquad \gamma_k = \alpha_k + \sum_{j=1}^{J} \phi_{jk}.$$

where $\psi(\cdot)$ is the digamma function.

Given the current values of the free parameters $(\gamma, \boldsymbol{\phi})$, we find (pseudo) MLE of $\boldsymbol{\lambda}$ in a closed form:

$$\lambda_{kj} \propto \sum_{i=1}^{N} \phi_{ijk} x_{ij},$$

where $i$ is the individual index. Since no closed form solution is available for the pseudo MLE of $\alpha$, we need to use an iterative method to maximize the lower bound with respect to $\alpha$. The gradient and the Hessian for the Newton–Raphson algorithm are as follows:

$$(19) \qquad \frac{\partial L}{\partial \alpha_k} = N\left(\Psi\left(\sum_{k=1}^{K} \alpha_k\right) - \Psi(\alpha_k)\right) + \sum_{i=1}^{N}\left(\Psi(\gamma_{ik}) - \Psi\left(\sum_{k=1}^{K} \gamma_{ik}\right)\right),$$

$$(20) \quad \frac{\partial L}{\partial \alpha_{k_1} \alpha_{k_2}} = N\left(\delta_{k_1 = k_2} \cdot \Psi'(\alpha_{k_1}) - \Psi'\left(\sum_{k_2=1}^{K} \alpha_{k_2}\right)\right).$$

Computations for the variational approximation are simpler and less time-consuming than for MCMC, but the quality of approximation depends on a specific functional form of the likelihood.

The C code for our implementation of the estimation algorithms for the GoM model is provided as part of supplemental material available at http://imstat.org/aoas/supplements.

## 5. Extended GoM mixture and its estimation.
Although there is no time dimension in the basic GoM model, the latent class representation essentially describes individuals as stochastic "movers." Here, individuals may move between extreme profiles when they respond to different items on the questionnaire. With this observation, it is natural to extend the GoM model to incorporate potential "stayers," or those individuals that provide item responses in a deterministic fashion, analogous to longitudinal mover-stayer models [Blumen, Kogan and Holland (1955)]. In the extended GoM mixture model, one compartment represents "movers" determined by the GoM part and other compartments represent different kinds of "stayers" determined by specific extreme profiles or by particular cells in the contingency table. The extended GoM model can also be seen as a combination of latent class and GoM mixture modeling analogous to the extended finite mixture model by Muthen and Shedden (1999).

For our analysis in this paper, we introduce one compartment of "stayers" for a specific cell in the table, and leave the question of choosing the number and nature of compartments open. Our choice of the "stayers" cell was motivated by two observations. First, in the functional disability data from the NLTCS, the cell that corresponds to the healthy people who report no disabilities is particularly difficult to fit with the standard GoM model. Thus, the excess of healthy people can be thought of as a set of outliers with respect to the standard GoM model. Second, it is known that elderly people move not only from being healthy to being disabled but also from being disabled to being healthy [Gill, Hardy and Williams (2002), Gill and Kurland (2003), Lynch, Brown and Harmsen (2003), Manton (1988)]. Therefore, even though the NLTCS participants are initially screened for chronic disability, it is reasonable to assume the presence of healthy "stayers" in the data. Accordingly, we assume that some proportion of people has zero probability to report a functional activity problem at the time of the survey and that everyone else has nonzero chances to report a functional disability problem according to the basic GoM model. Our specific example of extended GoM mixture model can also be thought of as analogous to zero-inflated Poisson regression [Lambert ([1992](#))].

Parameter estimation for the compartmental GoM model would be identical to the estimation for the standard GoM model if we knew how many individuals are healthy "stayers." Given that the number of healthy "stayers" is not observed, we need to modify parameter estimation taking into account a deterministic component.

More formally, we assume existence of: (1) a deterministic compartment of healthy individuals and (2) a stochastic GoM compartment. We denote by $\theta = (\theta_1, \theta_2)$ the respective weights such that $\theta_1 + \theta_2 = 1$. Assume individuals in the healthy compartment have no disabilities with probability 1. The distribution of responses for "movers" is given by the GoM model with parameters $\alpha, \lambda$.

We further augment the data with compartmental indicators to derive the following modifications for the MCMC sampling algorithm. Let $N$ be the total number of individuals in the sample and $n_2^{(m)}$ be the expected value of the all-zero cell count for the GoM compartment at the $m$th iteration. The expected value of the all-zero cell count from the healthy compartment, $n_1^{(m)}$, can be obtained by subtracting $n_2^{(m)}$ from the observed all-zero cell count. Denote the number of individuals with at least one positive and at least one zero response in their response pattern by $n_{\text{mix}}$. The total number of individuals from the GoM compartment at the $m$th iteration is then $n_2^{(m)} + n_{\text{mix}}$. We let the prior distribution for compartmental weights $\theta$ be uniform on the simplex, and update $\theta$ at the end of the posterior step with

$$(21) \qquad \theta_1^{(m+1)} = \theta_1^{(m)} + \frac{n_2^{(m)} - n_2^{(m+1)}}{N},$$

$$(22) \qquad \theta_2^{(m+1)} = \frac{n_2^{(m+1)} + n_{\text{mix}}}{N} = 1 - \theta_1^{(m+1)}.$$

We can easily generalize the algorithm to more than two compartments.

## 6. Model selection: Choice of dimensionality.

*Choice of dimensionality*: *Overview.* Statistical model selection methods include the Pearson's chi-square goodness of fit test [Pearson ([1900](#))], cross-validation techniques [Hastie, Tibshirani and Friedman ([2001](#))], penalized likelihood criteria such as the Akaike information criterion (AIC) [Akaike (1973)], the Bayesian Information Criterion (BIC) [Pelleg and Moore (2000), Schwartz (1978)] and Bayes factors [Kass and Raftery (1995)], reversible jump MCMC techniques [Green (1995)], deviance information criteria (DIC) [Spiegelhalter et al. (2002)] and more recent simulation-based analogues to AIC and BIC, called Akaike Information Criterion Monte Carlo (AICM) and Bayesian Information Criterion Monte Carlo (BICM) [Raftery et al. (2007)], among others.

Some of these criteria, AIC and BIC in particular, have been criticized as being not applicable for assessing the number of mixture components due to violations of the regularity conditions [McLachlan and Peel (2000)]. However, in spite of this, researchers continue to apply both criteria and to study their performance in a mixture context. Findings in population-level mixture models suggest that AIC tends to overestimate the correct number of components [Celeux and Soromenho (1996)], while BIC shows better results [Leroux (1992), Roeder and Wasserman (1997)].

Questions of dimensionality choice in mixed membership or individual-level mixture models have been approached by several authors [Airoldi et al. (2007), Blei, Ng and Jordan (2001), Erosheva (2002), Griffiths and Steyvers (2004), Pritchard, Stephens and Donnelly (2000)]. With one recent exception [Airoldi et al. (2007)], however, comparative performances of different selection criteria were not examined. Here, we provide an overview of several computationally feasible criteria and present results from a simulation study where we compare their performance in the context of the GoM model.

*Model selection criteria*: *Overview.* The Pearson chi-square test is one of the most common goodness-of-fit tests. It is not easily applicable to large sparse tables because of a large number of very small counts and, in the present context, because of the way in which the estimation is done, even if sparseness wasn't a problem, it wouldn't follow the usual chi-square distribution. We find it instructive, nonetheless, to examine deviations between expected and observed counts for cells with large observed values via the sum of squared Pearson residuals, see Bishop, Fienberg and Holland (1975). We refer to this criterion as the truncated sum of squared Pearson residuals (SSPR) criterion or $\chi^2_{\mathrm{tr}}$.

To calculate the truncated SSPR criterion, one needs to obtain expected values for selected response patterns $r = (r_1, \ldots, r_J)$, where $r_j = 0$ or $1$. For example, using draws from the posterior distribution, $\alpha_k^{(s)}$ and $\lambda_{kj}^{(s)}$, $s = 1, \ldots, S$, and a draw $g^{(s)} = (g_1^{(s)}, \ldots, g_K^{(s)})$ from $Dir(\alpha^{(s)})$, the expected count for response pattern $r$ can be computed as

$$\text{Expected Count} = \left( \frac{1}{S} \sum_{s=1}^{S} \prod_{j=1}^{J} \left( \sum_{k=1}^{K} g_k^{(s)} (\lambda_{kj}^{(s)})^{r_j} (1 - \lambda_{kj}^{(s)})^{1-r_j} \right) \right) \times N.$$

Note that label switching could present a problem for calculating posterior means and the model selection criteria based on them [Stephens (2000)].

For the variational approximation, the expected count for response pattern $r$ can be obtained as follows. Let $\hat{\alpha}$ and $\hat{\lambda}$ be the pseudo MLE obtained via variational approximation and let $g^{(s)}$, $s = 1, \ldots, S$, be draws from $Dir(\hat{\alpha})$, for some large $S$ (e.g., $S = 5000$). Then, the expected count for response pattern $r$ can be computed as above but with $\hat{\lambda}_{kj}$ in place of $\lambda_{kj}^{(s)}$ and with $g_k^{(s)}$ computed using $Dir(\hat{\alpha})$.

A general formulation of the BIC is based on the log-likelihood $\ell(x; \theta)$ and a maximum likelihood estimate $\hat{\theta}$:

$$BIC = -2\ell(x; \hat{\theta}) + p \log(N),$$

where $p$ is the number of free parameters in the model and $N$ is the number of data points. To obtain the BIC for the GoM model, we need to evaluate the integrated log-likelihood $\ell(x; \theta)$ at the maximum likelihood estimate of the parameter vector $\theta = (\lambda, \alpha)$. Since the GoM integrated likelihood is intractable, we use variational methods described in Section 4.1 to obtain an approximation to the BIC:

$$\widetilde{BIC} = -2\tilde{\ell}(x; \hat{\lambda}, \hat{\alpha}) + p \log(N),$$

where $\hat{\lambda}$ and $\hat{\alpha}$ are the (pseudo) maximum likelihood estimates obtained via variational approximation and $\tilde{\ell}(x; \hat{\lambda}, \hat{\alpha})$ is the lower bound on the log-likelihood. Models with larger values of $\widetilde{BIC}$ are preferable.

Bayesian measures of model complexity and fit, namely, DIC, AICM, and BICM, are analogous to AIC and BIC but are based solely on posterior simulation. While these criteria are attractive because of their computational simplicity for a given MCMC simulation, they may present other challenges such as choice of the parameters in focus [Spiegelhalter et al. (2002)].

A general formulation of DIC is based on the concepts of Bayesian deviance and the effective number of parameters. Bayesian deviance is defined as

$$D(\theta) = -2\ell(\theta) + 2\log(h(x)),$$

where $\ell(\theta) = \log p(x|\theta)$ and $h(x)$ is a function of the data only. Defining the effective number of parameters as

$$p_D = \overline{D(\theta)} - D(\overline{\theta}),$$

we compute DIC as follows:

$$DIC = D(\overline{\theta}) + 2p_D.$$

If we focus on GoM parameters $\theta = (\mathbf{g}, \boldsymbol{\lambda})$, we can compute a version of DIC directly using $S$ draws from the posterior distribution, $g_{ik}^{(s)}$ and $\lambda_{kj}^{(s)}$, $s = 1, \ldots, S$. The two pieces that we need to compute for DIC are

$$D(\overline{g}, \overline{\boldsymbol{\lambda}}) = -2 \sum_{i=1}^{N} \sum_{j=1}^{J} \log \left( \sum_{k=1}^{K} \overline{g_{ik}} \, \overline{\lambda_{kj}}^{x_{ij}} (1 - \overline{\lambda_{kj}})^{1-x_{ij}} \right),$$

where $\overline{g_{ik}} = \frac{1}{S} \sum_{s=1}^{S} g_{ik}^{(s)}$, and $\overline{\lambda_{kj}} = \frac{1}{S} \sum_{s=1}^{S} \lambda_{kj}^{(s)}$, and,

$$\overline{D(g, \boldsymbol{\lambda})} = -2 \frac{1}{S} \sum_{s=1}^{S} \sum_{i=1}^{N} \sum_{j=1}^{J} \log \left( \sum_{k=1}^{K} g_{ik}^{(s)} (\lambda_{kj}^{(s)})^{x_{ij}} (1 - \lambda_{kj}^{(s)})^{1-x_{ij}} \right).$$

Models with smaller values of DIC are preferable.

AICM is a penalized version of the posterior mean of the log-likelihoods

$$AICM = 2(\overline{\ell(\theta)} - s_{\ell(\theta)}^2),$$

that can be obtained using only the draws from the posterior simulation [Raftery et al. (2007)]. For the GoM model, the two pieces we need to compute are

$$\overline{\ell(\theta)} = \frac{1}{S} \sum_{s=1}^{S} \ell(\theta^{(s)}) \quad \text{and} \quad s_{\ell}^2 = \frac{1}{S} \sum_{s=1}^{S} (\ell(\theta^{(s)}) - \overline{\ell(\theta)})^2,$$

where $\theta = (\mathbf{g}, \boldsymbol{\lambda})$. Notice that $\ell(\theta) = -D(\theta)/2$.

**7. Simulation study.** We conducted a simulation study to investigate performance of the MCMC and variational approximation methods with respect to parameter recovery and dimensionality selection. Here, we briefly report main findings from this study.

We selected data generating designs to reflect several important features of functional disability data. Most noticeably, contingency tables on disability data often have a large number of zero or very small observed cell counts and several very large cell counts. Large cell counts typically include the all-zero and the all-one response patterns.

We considered 3- and 7-profile data generation scenarios. In the first scenario, we generated 5,000 individual responses on 16 binary items using the GoM model with $K^* = 3$ extreme profiles. We chose the profiles to be considered as "healthy," "disabled" and "intermediate" by their conditional response probabilities. Respective proportions of the profiles were $0.7, 0.2$ and $0.1$, and the hyperparameter was set at $\alpha_0 = 0.25$ to reflect the fact that individual responses to most items come from one extreme profile.

In the second scenario, we generated 5,000 individual responses to 10 binary items using the GoM model with 7 extreme profiles. We chose conditional response probabilities for 4 profiles so that they could be considered as "very healthy," "healthy," "disabled" and "very disabled." The other 3 intermediate profiles did not follow the ordering. Profile proportions ranged from 0.05 for one of the intermediate profiles to 0.4 for the "healthy" profile, and the hyperparameter was set at $\alpha_0 = 0.2$.

TABLE 1
*Choices of optimal K according to different model selection criteria*

| Criterion | Method | $K^* = 3$ | $K^* = 7$ |
|-----------|--------|-----------|-----------|
| $\chi^2_{\mathrm{tr}}$ | VA | **3** | **7** |
| $\chi^2_{\mathrm{tr}}$ | MCMC | 5 | **7** |
| BIC | VA | 2 | **7** |
| DIC | MCMC | 5 | 9 |
| AICM | MCMC | **3** | **7** |
| BICM | MCMC | 2 | 5 |

Under both scenarios, we carried out parameter estimation using the MCMC and variational approximation methods for the true values of $K^*$. The variational methods consistently provided better estimation for the observed count at the all-zero pattern, however, the MCMC approach yielded an overall better fit for the second scenario with $K^* = 7$. For $K^* = 3$, conditional probabilities for the "healthy" and "disabled" profiles were recovered very well with both estimation methods. For the intermediate profile, the variational approximation consistently overestimated and the MCMC consistently underestimated the conditional response probabilities. Parameter recovery was noticeably better for $K^* = 3$ than for $K^* = 7$. This could indicate that the number of items and the sample size in the second scenario were too small to provide reliable distinction among different grades of membership, given the selected extreme profiles and hyperparameter values. Variational estimates of the profiles' proportions were closer to the true values than corresponding MCMC estimates under both scenarios. In addition, the MCMC estimates of profile proportions had smaller range than the VA estimates.

For the MCMC method, given a sufficiently long run, the starting values of $\boldsymbol{\lambda}$ did not seem to influence the results, up to a relabeling of the extreme profiles. For smaller values of $K$, the posterior means obtained through MCMC simulations were very similar for all starting points considered. For this reason we did not use several starting points in higher dimensional cases that would have required substantial increases of computing time.

To investigate performance of different fit indices, we fitted the generated data sets using both the MCMC and variational methods separately for several values of $K$. For true $K^* = 3$, models fitted with $K = 2, 3, 4, 5$ were considered; for true $K^* = 7$, models fitted with $K = 5, 7, 9$ were considered. Table 1 summarizes results of six goodness-of-fit criteria, two of which rely on the variational approximation method, while the rest rely on the full MCMC calculations. The values of $\chi^2_{\text{tr}}$ were calculated for response patterns with observed counts $\geq 30$ in the first case and $\geq 40$ in the second case.

We see that $\chi^2_{\text{tr}}$ obtained with the variational method and AICM criteria perform well for both data generating scenarios, while BICM underestimates and DIC overestimates the true number of profiles in both cases. The variational approximation to BIC underestimates the model complexity for the 3-profile case, while it points to the true optimal number of profiles in the 7-profile case.

## 8. Analysis of the NLTCS functional disability data.

Data on 16 binary ADL and IADL items, pooled across four survey waves, 1982, 1984, 1989 and 1994, form a $2^{16}$ contingency table.[2] The total sample size is 21,574. Item marginal frequencies range from 0.1 for difficulty with eating to 0.7 for doing heavy housework. About 80% of cells in the contingency table have observed counts that are less than 5; 24 cells have observed counts greater than 100. These 24 most frequent response patterns account for 42% of the total observations (Table 2).

From an interpretative standpoint, it is often desirable to have data that satisfy latent unidimensionality, as in the well-known Rasch model. We formally tested the hypothesis of latent unidimensionality, following the approach of Holland and Rosenbaum (1986), using series of Mantel–Haenszel tests to detect negative conditional association among the 16 variables. We concluded that no monotone unidimensional latent structure model (e.g., one-factor or unidimensional logistic item response models) can provide an acceptable fit for the NLTCS data on 16 ADL/IADL items. Having rejected latent unidimensionality, our next step is to use the GoM analysis to determine characteristics and the number of disability profiles in the data.

The most apparent feature of the data is the very large observed count of "healthy" people. Almost 18% report no disabilities (Table 2), despite the fact that the majority of the NLTCS survey participants had been screened-in earlier as chronically disabled. A large fraction of "healthy" respondents includes disability recoveries, as well as survey supplements of the healthy and oldest-old in the 1994 wave. Since most of the "healthy" individuals have been identified earlier as chronically disabled, it is important to incorporate these responses in our model. We use the compartmental GoM model to estimate weights of deterministically healthy and partially disabled components. In doing so, we allow the extended GoM model to fit the observed count for the all-zero pattern. In addition, we examine the impact of the introduction of the "healthy" compartment on parameter estimates and on model choice.

---

[2]The full table is available for downloading from STATLIB at http://lib.stat.cmu.edu/ under the label *NLTCS*.

*Expected cell counts for 24 most frequent response patterns under the basic GoM model with K profiles*

| $n$ | Response pattern | Observed | Number of latent profiles $K$ | | | | | | | | |
|---|---|---|---|---|---|---|---|---|---|---|---|
| | | | 3 | 4 | 5 | 6 | 7 | 8 | 9 | 10 | 15 |
| 1 | 0 0 0 0 0 1 0 0 0 0 0 0 0 0 0 0 | 3853 | 2569 | 2055 | 2801 | 2889 | 3093 | 2941 | 3269 | 3016 | 3031 |
| 2 | 0 0 0 0 1 0 0 0 0 0 0 0 0 0 0 0 | 216 | 225 | 172 | 177 | 186 | 180 | 180 | 202 | 205 | 187 |
| 3 | 0 0 0 0 0 0 1 0 0 0 0 0 0 0 0 0 | 1107 | 1135 | 710 | 912 | 993 | 914 | 937 | 1010 | 944 | 940 |
| 4 | 0 0 0 0 1 0 1 0 0 0 0 0 0 0 0 0 | 188 | 116 | 76 | 113 | 200 | 199 | 181 | 190 | 198 | 201 |
| 5 | 0 0 0 0 0 0 1 0 0 0 1 0 0 0 0 0 | 122 | 64 | 88 | 58 | 199 | 90 | 89 | 116 | 127 | 127 |
| 6 | 0 0 0 0 0 0 0 0 0 0 0 1 0 0 0 0 | 351 | 344 | 245 | 250 | 274 | 274 | 259 | 331 | 303 | 357 |
| 7 | 0 0 1 0 0 0 0 0 0 0 0 1 0 0 0 0 | 206 | 20 | 23 | 116 | 86 | 80 | 137 | 116 | 111 | 149 |
| 8 | 0 0 0 0 0 0 1 0 0 0 1 0 0 0 0 0 | 303 | 200 | 126 | 324 | 255 | 236 | 213 | 273 | 264 | 325 |
| 9 | 0 0 1 0 0 0 1 0 0 0 0 1 0 0 0 0 | 182 | 44 | 71 | 170 | 169 | 162 | 200 | 172 | 187 | 219 |
| 10 | 0 0 0 0 1 0 1 0 0 0 0 1 0 0 0 0 | 108 | 51 | 39 | 162 | 105 | 85 | 117 | 97 | 108 | 116 |
| 11 | 0 0 1 0 1 0 1 0 0 0 0 1 0 0 0 0 | 106 | 32 | 94 | 94 | 123 | 125 | 133 | 142 | 157 | 136 |
| 12 | 0 0 0 0 1 0 0 0 0 0 0 0 1 0 0 0 | 195 | 219 | 101 | 160 | 46 | 25 | 24 | 25 | 31 | 27 |
| 13 | 0 0 0 0 0 0 1 0 0 0 0 0 1 0 0 0 | 198 | 127 | 111 | 108 | 341 | 170 | 169 | 189 | 200 | 163 |
| 14 | 0 0 0 0 0 0 1 0 0 0 1 0 1 0 0 0 | 196 | 41 | 172 | 90 | 104 | 224 | 214 | 174 | 187 | 160 |
| 15 | 0 0 0 0 0 0 1 0 0 0 1 1 0 0 0 0 | 123 | 96 | 86 | 132 | 131 | 120 | 109 | 95 | 108 | 110 |
| 16 | 0 0 0 0 0 0 1 0 0 0 1 1 1 0 0 0 | 176 | 136 | 162 | 97 | 67 | 167 | 149 | 152 | 167 | 157 |
| 17 | 0 0 1 0 0 0 1 0 0 0 1 1 1 0 0 0 | 120 | 144 | 104 | 41 | 57 | 47 | 96 | 75 | 72 | 80 |
| 18 | 0 0 0 0 1 0 1 0 0 0 1 1 1 0 0 0 | 101 | 127 | 90 | 54 | 41 | 68 | 72 | 70 | 74 | 124 |
| 19 | 0 1 1 1 1 1 1 1 1 1 1 1 1 0 0 0 | 102 | 44 | 38 | 22 | 18 | 18 | 85 | 103 | 85 | 61 |
| 20 | 1 1 1 1 1 1 1 1 1 1 1 1 0 1 0 | 107 | 88 | 104 | 96 | 84 | 87 | 43 | 37 | 31 | 73 |
| 21 | 0 1 1 1 1 1 1 1 1 1 1 1 0 1 0 | 104 | 269 | 239 | 202 | 52 | 50 | 50 | 63 | 53 | 66 |
| 22 | 1 1 1 1 1 1 1 1 1 1 1 1 1 1 0 | 164 | 214 | 246 | 272 | 274 | 276 | 224 | 166 | 143 | 115 |
| 23 | 0 1 1 1 1 1 1 1 1 1 1 1 1 1 1 | 153 | 291 | 261 | 266 | 250 | 230 | 235 | 189 | 167 | 137 |
| 24 | 1 1 1 1 1 1 1 1 1 1 1 1 1 1 1 | 660 | 233 | 270 | 362 | 419 | 418 | 582 | 612 | 474 | 423 |

### 8.1. *GoM analysis.*

*MCMC sampling.* We applied the fully Bayesian approach from Section 4 to estimate the posterior distribution of the GoM model parameters with the number of extreme profiles ranging from $K = 3$ to $K = 15$. Extreme profiles for $K = 2$ were identified as "healthy" and "disabled," making the $K = 2$ GoM model a monotone unidimensional latent structure model. Having rejected latent unidimensionality earlier, we only considered results for $K = 3$ and beyond in the rest of our analysis.

We expected individual vectors of membership scores to be dominated by one component, hence, we set the prior for $\alpha_0$ to be Gamma(2, 10). We chose the prior for the relative proportions $\xi$ to be uniform on the simplex and put uniform independent priors on the conditional response probabilities $\boldsymbol{\lambda}$.

We fit the models sequentially in the order of $K$. For the GoM model with $K$ extreme profiles, we set starting values for $\boldsymbol{\lambda}$ to the estimated conditional response probabilities from the latent class model with $K$ classes. We took the posterior mean of $\alpha_0$ from the GoM model with $K - 1$ extreme profiles to be the starting value for $\alpha_0$ for the GoM model with $K$ extreme profiles. We chose starting values for the hyperparameters $\xi$ to be equal to the latent class weights estimated from the $K$ class model.

For each value of $K$, we adjusted the tuning parameters $\omega$ (for $\alpha_0$) and $\delta$ (for $\xi$) to reach a compromise between the acceptance rates of the Metropolis–Hastings steps and the amount of mixing. The acceptance rates for $\alpha_0$ and $\xi$ varied respectively from 11% and 28% in lower dimensions to 5% and 9% in higher dimensions. Since the acceptance rates were low, we introduced thinning parameter $q$ and kept every $q$th draw and discarded the rest; $q$ varied from 10 in lower dimensions to 140 in higher dimensions.

Choosing the length of a burn-in period did not appear to be a problem with our data. The chains generally did not experience long burn-in periods, except when starting values for the hyperparameters were very far from the posterior means. The burn-in period varied from 10,000 iterations in lower dimensions to 60,000 in higher dimensions.

For each parameter, we monitored univariate convergence via Geweke diagnostics, and Heidelberger and Welch stationarity and interval halfwidth tests, available from the CODA package [Best, Cowles and Vines (1996)]. In addition, we visually examined plots of successive iterations. To assess convergence of the multivariate posterior, we examined successive values of the log-likelihood with the same set of methods. The chains needed far fewer iterations to converge in posterior means than they needed to converge in distribution for all parameters and the log-likelihood.

We ran all chains long enough to reach acceptable convergence levels. We had to consider larger number of iterations for higher values of $K$ to accommodate slow convergence of the hyperparameters due to slow mixing of the chains. The additional iterations needed to satisfy convergence criteria for hyperparameters (after the other parameters have reached convergence) had negligible effect on the posterior means of the conditional response probabilities.

Table 3
*Truncated sum of squared Pearson residuals, $\chi^2_{tr}$, for the basic GoM model with $K$ profiles, with different levels of truncation*

| | Number of latent profiles $K$ | | | | | | | | |
|---|---|---|---|---|---|---|---|---|---|
| Level | 3 | 4 | 5 | 6 | 7 | 8 | 9 | 10 | 15 |
| $\geq 100$ | 4889 | 5032 | 1840 | 2202 | 2458 | 1908 | 1582 | 1602 | 1604 |
| $\geq 25$ | 14562 | 10458 | 6153 | 4337 | 3566 | 2194 | 1803 | 1997 | 1946 |
| $\geq 10$ | 52288 | 20625 | 10839 | 7766 | 6251 | 4534 | 3931 | 4276 | 4258 |

*Model selection.* Table 2 provides 24 response patterns with observed cell counts $\geq 100$ and corresponding expected counts obtained using draws from the posterior distribution for each $K = 3, \ldots, 9, 10, 15$. We observe that the model with $K = 9$ replicated the marginal pattern abundance best. It is especially evident that models with $K = 10$ and $K = 15$ did not fit the three largest cell counts as well as the 9-profile model.

To select the number of profiles, we used all of the criteria that performed well in our simulation study described in Section 6 (the truncated SSPR criterion, the variational approximation to the BIC, and the AICM). We also calculated DIC for a further comparison, although it overestimated the correct number of profiles in the simulation study.

Table 3 gives values of the truncated SSPR criterion, $\chi^2_{tr}$, for three different levels of truncation, over cells with observed counts $\geq 100, 25$ and 10. All three criteria indicate that the $K = 9$ model has a better fit in an absolute sense, that is, without correcting for differences in the degrees of freedom.

Figure 1 shows plots of the DIC, the BIC approximation, the AICM and the truncated SSPR criterion for the 100 level of truncation. All criteria agree that the optimal number of profiles is greater than 7. Recall that in the 7-profile simulation study the AICM, the BIC approximation, and the truncated SSPR criterion all obtained the correct number of components. For the NLTCS data, these criteria point to 7, 10 and 9 profiles, respectively. Although the DIC overestimated the correct number of profiles in our simulation study, it indicates that 9 profiles is the optimal number for the NLTCS functional disability data. The value of $K = 9$ is in agreement with the results from truncated SSPR, but is less than the optimal choice of $K = 10$ identified by BIC.

We used the following steps to verify that no label switching had occurred in the MCMC chains. First, we postulated that label switching occasions should be visible as jumps in trace plots of the MCMC iterations when the extreme profiles are well separated in the multidimensional space. We found extreme profiles to be well separated in the multidimensional space for all $K < 9$. That is, there was at least one item for which posterior means were at least two standard deviations away from each other for each pair of the profiles (Table 4). We then visually monitored chains to identify jumps that could correspond to label permutations in the posterior distribution. We observed no jumps for models with $K < 9$ and concluded that no label switching occurred in those chains.

We weren't able to carry out analogous assessments for the GoM models with $K = 9$ and higher since the profiles were no longer well separated (compare, e.g., profiles $k = 1$ and 8 in Table 4.[3]) It is possible that label-switching did occur in those cases which would question validity of posterior mean estimates and the use of DIC and AICM. However, given that the approximate BIC, which is not impacted by label switching, indicated $K^* = 10$, a choice of an optimal $K$ around that value seems reasonable.

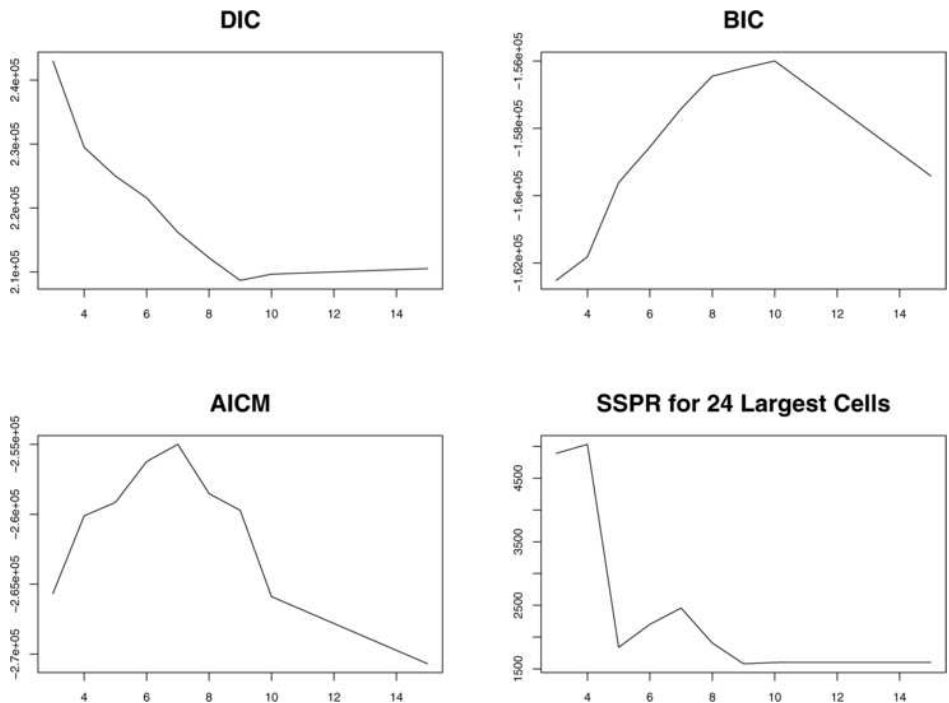

Fig. 1. *Goodness-of-fit criteria for the basic GoM model.*

---

[3]Standard deviation estimates are provided as part of supplemental material available at http://imstat.org/aoas/supplements.

Table 4
*Posterior mean estimates for the basic GoM model with 9 profiles*

| | Extreme profile number ($k$) | | | | | | | | |
|---|---|---|---|---|---|---|---|---|---|
| | **1** | **2** | **3** | **4** | **5** | **6** | **7** | **8** | **9** |
| $\hat{\lambda}_{k,1}$ | 0.001 | 0.035 | 0.002 | 0.005 | 0.239 | 0.002 | 0.738 | 0.001 | 0.002 |
| $\hat{\lambda}_{k,2}$ | 0.001 | 0.071 | 0.003 | 0.269 | 0.891 | 0.437 | 0.967 | 0.001 | 0.001 |
| $\hat{\lambda}_{k,3}$ | 0.001 | 0.285 | 0.001 | 0.706 | 0.994 | 0.875 | 0.976 | 0.001 | 0.004 |
| $\hat{\lambda}_{k,4}$ | 0.009 | 0.158 | 0.029 | 0.076 | 0.674 | 0.080 | 0.970 | 0.004 | 0.013 |
| $\hat{\lambda}_{k,5}$ | 0.070 | 0.550 | 0.171 | 0.453 | 0.974 | 0.627 | 0.998 | 0.039 | 0.266 |
| $\hat{\lambda}_{k,6}$ | 0.011 | 0.114 | 0.026 | 0.208 | 0.774 | 0.317 | 0.894 | 0.005 | 0.026 |
| $\hat{\lambda}_{k,7}$ | 0.008 | 0.985 | 0.973 | 0.607 | 0.999 | 0.948 | 0.999 | 0.007 | 0.761 |
| $\hat{\lambda}_{k,8}$ | $< 0.001$ | 0.524 | 0.019 | 0.005 | 0.669 | 0.034 | 0.955 | $< 0.001$ | 0.011 |
| $\hat{\lambda}_{k,9}$ | 0.001 | 0.909 | 0.093 | 0.034 | 0.864 | 0.412 | 0.997 | 0.001 | 0.208 |
| $\hat{\lambda}_{k,10}$ | 0.001 | 0.822 | 0.014 | 0.001 | 0.694 | 0.067 | 0.998 | 0.001 | 0.055 |
| $\hat{\lambda}_{k,11}$ | 0.002 | 0.977 | 0.080 | 0.077 | 0.920 | 0.856 | 0.995 | 0.002 | 0.752 |
| $\hat{\lambda}_{k,12}$ | 0.042 | 0.692 | 0.146 | 0.933 | 0.950 | 0.998 | 0.936 | 0.076 | 0.448 |
| $\hat{\lambda}_{k,13}$ | 0.037 | 0.836 | 0.109 | 0.219 | 0.838 | 0.847 | 0.894 | 0.037 | 0.849 |
| $\hat{\lambda}_{k,14}$ | 0.012 | 0.626 | 0.013 | 0.002 | 0.230 | 0.144 | 0.908 | 0.007 | 0.282 |
| $\hat{\lambda}_{k,15}$ | 0.022 | 0.489 | 0.055 | 0.029 | 0.345 | 0.068 | 0.909 | 0.010 | 0.127 |
| $\hat{\lambda}_{k,16}$ | 0.024 | 0.386 | 0.021 | 0.007 | 0.061 | 0.027 | 0.768 | 0.017 | 0.099 |
| $\hat{\xi}_k$ | 0.095 | 0.107 | 0.111 | 0.114 | 0.115 | 0.114 | 0.114 | 0.114 | 0.114 |
| $\hat{\alpha}_0$ | 0.095 | | | | | | | | |

ADL items: (1) eating, (2) getting in/out of bed, (3) getting around inside, (4) dressing, (5) bathing, (6) using toilet. IADL items: (7) doing heavy housework, (8) doing light housework, (9) doing laundry, (10) cooking, (11) grocery shopping, (12) getting about outside, (13) traveling, (14) managing money, (15) taking medicine, (16) telephoning.

We examined the estimated profiles for $K = 7$ and $K = 9$ GoM models. Contrary to our expectations, we did not find the interpretation of the 7-profile model to be more appealing from a substantive point of view. Therefore, we report the estimated profiles for the 9-profile GoM model that is identified as the optimal by truncated SSPR criteria. Table 4 provides posterior means and standard deviations for the conditional response probabilities, $\lambda_{kj} = \mathrm{pr}(x_j = 1 | g_k = 1)$; these are probabilities of being disabled on activity $j$ for a complete member of extreme profile $k$. Estimation via variational methods yielded similar results in terms of profile interpretation, although variational estimates of conditional probabilities were generally closer to the boundaries of the parameter space.

Given that the fit of the all-zero pattern is still not very good for the 9-profile GoM model, we turn to the extended GoM mixture model, incorporating a "deterministically" healthy compartment.

TABLE 5

*Expected cell counts for 23 most frequent response patterns under extended GoM mixture model with K profiles and a healthy compartment*

| | | Number of latent profiles $K$ | | | | | | | | |
|---|---|---|---|---|---|---|---|---|---|---|
| $n$ | Response pattern | Observed | 3 | 4 | 5 | 6 | 7 | 8 | 9 | 10 |
| 2 | 0 0 0 0 1 0 0 0 0 0 0 0 0 0 0 | 216 | 77 | 133 | 139 | 151 | 136 | 152 | 201 | 148 |
| 3 | 0 0 0 0 0 0 1 0 0 0 0 0 0 0 0 | 1107 | 587 | 661 | 835 | 856 | 799 | 897 | 933 | 845 |
| 4 | 0 0 0 0 1 0 1 0 0 0 0 0 0 0 0 | 188 | 142 | 162 | 203 | 204 | 197 | 194 | 258 | 167 |
| 5 | 0 0 0 0 0 0 1 0 0 0 1 0 0 0 0 | 122 | 209 | 59 | 118 | 84 | 86 | 113 | 173 | 97 |
| 6 | 0 0 0 0 0 0 0 0 0 0 0 1 0 0 0 | 351 | 117 | 195 | 170 | 200 | 213 | 225 | 279 | 212 |
| 7 | 0 0 1 0 0 0 0 0 0 0 0 1 0 0 0 | 206 | 14 | 21 | 143 | 184 | 176 | 125 | 94 | 150 |
| 8 | 0 0 0 0 0 0 1 0 0 0 0 1 0 0 0 | 303 | 229 | 247 | 253 | 260 | 246 | 255 | 310 | 236 |
| 9 | 0 0 1 0 0 0 1 0 0 0 0 1 0 0 0 | 182 | 56 | 63 | 213 | 230 | 192 | 195 | 156 | 197 |
| 10 | 0 0 0 0 1 0 1 0 0 0 0 1 0 0 0 | 108 | 75 | 73 | 122 | 122 | 114 | 98 | 113 | 87 |
| 11 | 0 0 1 0 1 0 1 0 0 0 0 1 0 0 0 | 106 | 56 | 76 | 119 | 120 | 106 | 158 | 125 | 102 |
| 12 | 0 0 0 0 1 0 0 0 0 0 0 0 1 0 0 | 195 | 38 | 26 | 38 | 29 | 26 | 36 | 33 | 31 |
| 13 | 0 0 0 0 0 0 1 0 0 0 0 0 1 0 0 | 198 | 287 | 139 | 222 | 183 | 183 | 177 | 244 | 148 |
| 14 | 0 0 0 0 0 0 1 0 0 0 1 0 1 0 0 | 196 | 107 | 106 | 71 | 191 | 193 | 188 | 164 | 190 |
| 15 | 0 0 0 0 0 0 1 0 0 0 0 1 1 0 0 | 123 | 138 | 76 | 115 | 117 | 105 | 96 | 98 | 116 |
| 16 | 0 0 0 0 0 0 1 0 0 0 1 1 1 0 0 | 176 | 85 | 121 | 75 | 166 | 142 | 160 | 149 | 169 |
| 17 | 0 0 1 0 0 0 1 0 0 0 1 1 1 0 0 | 120 | 119 | 86 | 44 | 38 | 76 | 79 | 66 | 96 |
| 18 | 0 0 0 0 1 0 1 0 0 0 1 1 1 0 0 | 101 | 65 | 70 | 45 | 68 | 60 | 73 | 74 | 98 |
| 19 | 0 1 1 1 1 1 1 1 1 1 1 1 1 0 0 0 | 102 | 33 | 32 | 15 | 17 | 89 | 108 | 106 | 111 |
| 20 | 1 1 1 1 1 1 1 1 1 1 1 1 1 0 1 0 | 107 | 74 | 99 | 64 | 99 | 41 | 40 | 39 | 40 |
| 21 | 0 1 1 1 1 1 1 1 1 1 1 1 0 1 0 | 104 | 89 | 90 | 39 | 57 | 52 | 66 | 71 | 92 |
| 22 | 1 1 1 1 1 1 1 1 1 1 1 1 1 1 1 0 | 164 | 198 | 243 | 218 | 269 | 222 | 189 | 155 | 148 |
| 23 | 0 1 1 1 1 1 1 1 1 1 1 1 1 1 1 1 | 153 | 262 | 240 | 222 | 200 | 241 | 212 | 192 | 172 |
| 24 | 1 1 1 1 1 1 1 1 1 1 1 1 1 1 1 1 | 660 | 223 | 272 | 346 | 359 | 610 | 581 | 556 | 564 |

### 8.2. *The extended GoM mixture analysis.*

*MCMC sampling.* We carried out the extended GoM mixture analysis for $K = 3, \ldots, 9, 10$, as described in Section 5. We chose initial values, ran MCMC samplers, and determined convergence similarly as in Section 8.1. We set an initial value for the weight of the healthy compartment $\theta_1$ to be a positive fraction that is less than the observed proportion of individuals with all-zero responses.

*Model selection.* Table 5 provides the expected and observed cell counts for the 23 most frequent response patterns; we excluded the all-zero pattern since the extended GoM mixture fits it precisely. It is difficult to choose among $K = 7, 8$ or 9 based on the expected counts in Table 5, but the model for $K = 8$ shows the best fit as indicated by truncated SSPR over the differing levels of truncation in Table 6.

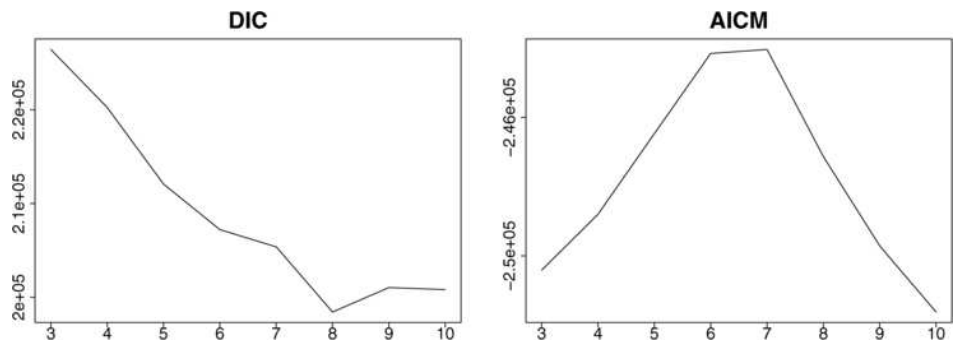

Fig. 2. *DIC (left) and AICM (right) for the GoM mixture model.*

Analogously to the standard GoM model, we computed a version of AICM and a version of DIC obtained directly from the MCMC output. The AICM plot in Figure 2 picks $K^* = 7$ profiles, while the DIC plot in Figure 2 suggests the choice of $K^* = 8$ profiles for the extended GoM mixture model, which is consistent with the SSPR selection. We therefore examine the 8-profile extended GoM mixture model.

Table 7 provides the conditional response probabilities for the 8 profiles that we interpret in detail at the end of this section.[4] Similarly to results from the standard GoM model with 9 profiles, estimated profile weights $\hat{\xi}_k$, $k = 1, \ldots, K$, are all close to $1/K$. Estimated proportions of the healthy compartment for $K = 3, \ldots, 10$ range from 14% to around 16% (Table 8). The estimated proportion of deterministically healthy individuals from the 8-profile GoM mixture model is $\hat{\theta}_1 = 0.15$ with the standard error of 0.006.

*Comparison of results for the basic and the extended GoM mixture models.* The optimal dimensionality values identified by truncated SSPR, AICM and DIC criteria for the extended GoM mixture model are one less than the

TABLE 6

*Truncated sum of squared Pearson residuals, $\chi^2_{\text{tr}}$, for extended GoM mixture models with $K$ profiles and a healthy compartment, with different levels of truncation*

| | Number of latent profiles $K$ | | | | | | | |
|---|---|---|---|---|---|---|---|---|
| **Level** | **3** | **4** | **5** | **6** | **7** | **8** | **9** | **10** |
| $\geq 100$ | 6169 | 4493 | 2541 | 2171 | 1666 | 1139 | 1265 | 1285 |
| $\geq 25$ | 13611 | 8689 | 4605 | 4246 | 2582 | 1739 | 2211 | 2276 |
| $\geq 10$ | 24638 | 14736 | 9120 | 6647 | 4678 | 3738 | 4028 | 4215 |

---

TABLE 7
*Posterior mean estimates for the extended GoM mixture model with 8 extreme profiles
and a healthy compartment*

| | Extreme profile number ($k$) | | | | | | | |
|---|---|---|---|---|---|---|---|---|
| | **1** | **2** | **3** | **4** | **5** | **6** | **7** | **8** |
| $\hat{\lambda}_{k,1}$ | 0.004 | 0.243 | 0.002 | 0.008 | 0.002 | 0.740 | 0.034 | 0.002 |
| $\hat{\lambda}_{k,2}$ | 0.005 | 0.900 | 0.448 | 0.288 | 0.003 | 0.970 | 0.079 | 0.001 |
| $\hat{\lambda}_{k,3}$ | 0.003 | 0.996 | 0.889 | 0.742 | 0.001 | 0.978 | 0.296 | 0.005 |
| $\hat{\lambda}_{k,4}$ | 0.025 | 0.685 | 0.083 | 0.081 | 0.029 | 0.972 | 0.158 | 0.013 |
| $\hat{\lambda}_{k,5}$ | 0.196 | 0.978 | 0.634 | 0.445 | 0.165 | 0.998 | 0.554 | 0.263 |
| $\hat{\lambda}_{k,6}$ | 0.039 | 0.783 | 0.327 | 0.212 | 0.024 | 0.897 | 0.116 | 0.024 |
| $\hat{\lambda}_{k,7}$ | 0.101 | 0.999 | 0.946 | 0.604 | 0.938 | 0.999 | 0.982 | 0.772 |
| $\hat{\lambda}_{k,8}$ | 0.001 | 0.686 | 0.032 | 0.005 | 0.017 | 0.956 | 0.525 | 0.013 |
| $\hat{\lambda}_{k,9}$ | 0.002 | 0.873 | 0.412 | 0.036 | 0.088 | 0.998 | 0.908 | 0.221 |
| $\hat{\lambda}_{k,10}$ | 0.002 | 0.705 | 0.065 | 0.001 | 0.014 | 0.998 | 0.820 | 0.057 |
| $\hat{\lambda}_{k,11}$ | 0.038 | 0.923 | 0.858 | 0.081 | 0.067 | 0.995 | 0.975 | 0.769 |
| $\hat{\lambda}_{k,12}$ | 0.234 | 0.949 | 0.998 | 0.916 | 0.146 | 0.934 | 0.697 | 0.444 |
| $\hat{\lambda}_{k,13}$ | 0.180 | 0.838 | 0.853 | 0.212 | 0.095 | 0.892 | 0.833 | 0.834 |
| $\hat{\lambda}_{k,14}$ | 0.046 | 0.222 | 0.141 | 0.002 | 0.010 | 0.909 | 0.619 | 0.278 |
| $\hat{\lambda}_{k,15}$ | 0.057 | 0.343 | 0.066 | 0.029 | 0.053 | 0.909 | 0.484 | 0.122 |
| $\hat{\lambda}_{k,16}$ | 0.066 | 0.054 | 0.025 | 0.005 | 0.020 | 0.768 | 0.379 | 0.092 |
| $\hat{\xi}$ | 0.104 | 0.120 | 0.126 | 0.129 | 0.130 | 0.130 | 0.130 | 0.130 |
| $\hat{\alpha}_0$ | 0.103 | | | | | | | |
| $\hat{\theta}_1$ | 0.146 | | | | | | | |

ADL items: (1) eating, (2) getting in/out of bed, (3) getting around inside, (4) dressing,
(5) bathing, (6) using toilet. IADL items: (7) doing heavy housework, (8) doing light
housework, (9) doing laundry, (10) cooking, (11) grocery shopping, (12) getting about
outside, (13) traveling, (14) managing money, (15) taking medicine, (16) telephoning.

corresponding optimal values for the basic GoM model. The presence of the
deterministic healthy compartment therefore reduces the optimal number of
profiles by one in the NLTCS disability data.

The preferred dimensionality choices are $K^* = 9$ and $K^* = 8$ for the ba-
sic GoM and extended GoM mixture models, respectively. Comparing DIC
values for these models, we observe that the extended GoM mixture model
provides an improved fit to the data. Comparing the estimated conditional
response probabilities, we observe that all but two "healthy" profiles from
the 9-profile basic GoM model match seven estimated profiles from the 8-
profile GoM mixture model closely (see $k = 6$ in Table 7 and $k = 7$ in Ta-
ble 4, e.g.). Moreover, the two "healthy" profiles from the 9-profile basic
GoM model do not differ by much (see $k = 1$ and $k = 8$ in Table 4); in fact,

taking standard errors into account, they are identical.[5] The unmatched profile from the 8-profile GoM mixture model ($k = 1$ in Table 7) is the new healthy profile.

To aid interpretation, we compare the profiles' estimated conditional response probabilities to the average probabilities for each functional disability item. We would like to see by how much the frequency of disability occurrence for each profile differs from the average frequency of occurrence of the same functional disability in the population as a whole. Relative frequencies for profile $k$, obtained as

$$(23) \qquad \lambda_{kj}/\lambda_j, \qquad j = 1, \ldots, 16,$$

where $\lambda_j$ is the marginal probability for item $j$, indicate how frequently each disability is observed for a complete member of the extreme profile in relation to the population average (Table 9). For example, a complete member of extreme profile 6 is about seven times more likely to need help with eating than individuals in the NLTCS sample need on average (11%).

Table 9 shows values for relative probabilities greater than 1 in red ink. Among estimated extreme profiles in the 8-profile GoM mixture model, we find one healthy profile ($k = 1$) with all relative frequencies less than the corresponding population averages, while all other profiles have at least one activity with relative frequency greater than the population average. In addition, we find that each estimated profile in the 8-profile GoM mixture model has a unique set of functional disabilities with relative frequencies greater than the corresponding population averages (no two rows in the table have identical placements of values in red ink). We can say then that the estimated 8-profile GoM mixture solution defines a set of *admissible* profiles in the terminology of Berkman, Singer and Manton (1989). Moreover, taking into account standard errors of the estimates (not shown), we notice that all 8 disability profiles are now well separated.

TABLE 8
*Posterior mean estimates and standard deviations of the proportion in the healthy compartment for the extended GoM mixture model with K profiles*

| $K$ | 3 | 4 | 5 | 6 | 7 | 8 | 9 | 10 |
|---|---|---|---|---|---|---|---|---|
| $\hat{\theta}_1$ | 0.162 | 0.159 | 0.152 | 0.148 | 0.152 | 0.146 | 0.141 | 0.154 |
| SD($\hat{\theta}_1$) | 0.0024 | 0.004 | 0.0045 | 0.0054 | 0.0047 | 0.0061 | 0.006 | 0.0045 |

---

[5]MCMC estimation in our simulation studies identified emerging identical, up to a standard error, profiles when the number of fitted profiles was greater than the number of profiles that generated the data.

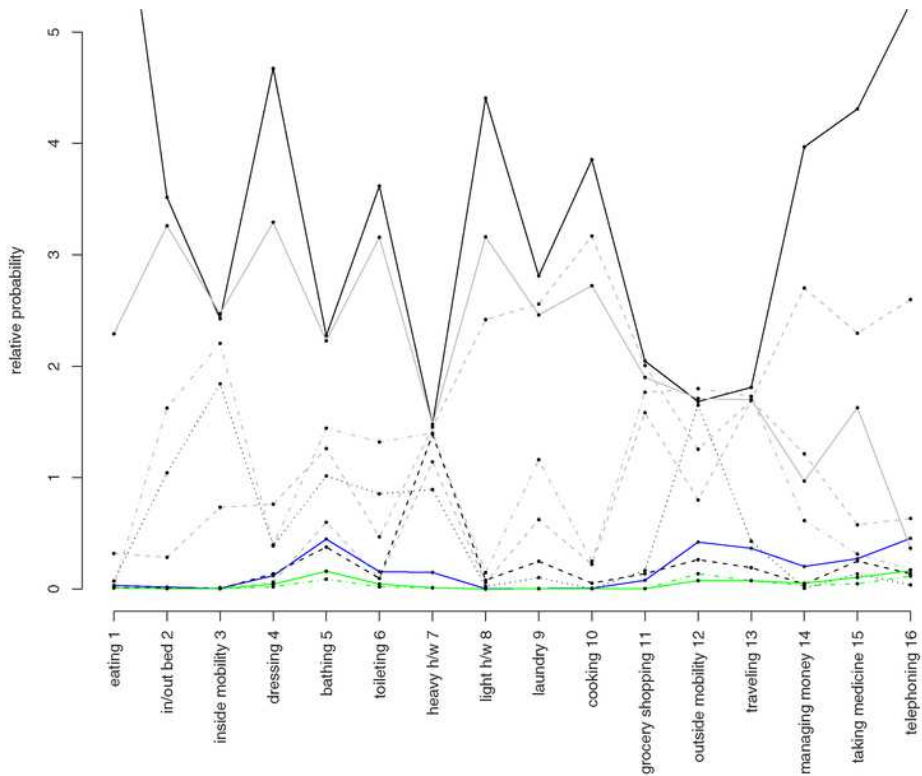

| Activities | eating 1 | in/out bed 2 | inside mobility 3 | dressing 4 | bathing 5 | toileting 6 | heavy h/w 7 | light h/w 8 | laundry 9 | cooking 10 | grocery shopping 11 | outside mobility 12 | traveling 13 | managing money 14 | taking medicine 15 | telephoning 16 |
|---|---|---|---|---|---|---|---|---|---|---|---|---|---|---|---|---|
| profile 1 | 0.03 | 0.02 | 0.01 | 0.12 | 0.45 | 0.16 | 0.15 | 0.00 | 0.00 | 0.01 | 0.08 | 0.42 | 0.37 | 0.20 | 0.27 | 0.46 |
| profile 2 | 2.29 | 3.26 | 2.47 | 3.29 | 2.23 | 3.16 | 1.48 | 3.16 | 2.46 | 2.72 | 1.90 | 1.71 | 1.70 | 0.70 | 1.63 | 0.37 |
| profile 3 | 0.02 | 1.62 | 2.21 | 0.40 | 1.44 | 1.32 | 1.40 | 0.15 | 1.16 | 0.25 | 1.77 | 1.78 | 1.73 | 0.62 | 0.32 | 0.17 |
| profile 4 | 0.07 | 1.04 | 1.84 | 0.39 | 1.02 | 0.85 | 0.80 | 0.02 | 0.10 | 0.00 | 0.17 | 1.65 | 0.43 | 0.01 | 0.14 | 0.04 |
| profile 5 | 0.01 | 0.01 | 0.00 | 0.14 | 0.38 | 0.10 | 1.39 | 0.08 | 0.25 | 0.05 | 0.14 | 0.26 | 0.19 | 0.04 | 0.25 | 0.14 |
| profile 6 | 6.98 | 3.52 | 2.43 | 4.67 | 2.27 | 3.62 | 1.48 | 4.41 | 2.81 | 3.85 | 2.05 | 1.68 | 1.81 | 3.97 | 4.31 | 5.26 |
| profile 7 | 0.32 | 0.29 | 0.73 | 0.76 | 1.26 | 0.47 | 1.45 | 2.42 | 2.56 | 3.17 | 2.01 | 1.26 | 1.69 | 2.70 | 2.30 | 2.60 |
| profile 8 | 0.02 | 0.00 | 0.01 | 0.06 | 0.60 | 0.10 | 1.14 | 0.06 | 0.62 | 0.22 | 1.58 | 0.80 | 1.69 | 1.21 | 0.58 | 0.63 |
| profile 1* | 0.01 | 0.00 | 0.00 | 0.04 | 0.16 | 0.04 | 0.01 | 0.00 | 0.00 | 0.00 | 0.00 | 0.08 | 0.07 | 0.05 | 0.1 | 0.16 |
| profile 8** | 0.01 | 0.00 | 0.00 | 0.02 | 0.09 | 0.02 | 0.01 | 0.00 | 0.00 | 0.00 | 0.00 | 0.14 | 0.07 | 0.03 | 0.05 | 0.12 |
| average frequency | 0.11 | 0.28 | 0.40 | 0.21 | 0.44 | 0.25 | 0.68 | 0.22 | 0.35 | 0.26 | 0.49 | 0.56 | 0.49 | 0.23 | 0.21 | 0.15 |

Table 9

*Functional disabilities average frequencies and relative frequencies by profile for the $K = 8$ GoM mixture model and for two healthy profiles from the basic $K = 9$ GoM model (green labels). Relative frequencies greater than 1 are in red*

Extreme profiles in black in Table 9 are the seven profiles from the 8-profile GoM mixture model that match corresponding profiles from the 9-profile basic GoM model closely. The two differing healthy profiles from the basic 9-profile GoM model are shown in green ink, and the new healthy profile from the GoM mixture model is in blue.

While Table 9 allows us to view all estimated profiles in relation to one other, it is not easy to trace each profile separately on this plot. Pairwise plots in Figure 3 allow us to view individual profiles in detail. Profile $k = 2$ exhibits relative conditional probabilities greater than 1 for all activities except managing money and telephoning. Relative probabilities for transferring in/out of bed, dressing, toileting, and light housework for this profile are at least three times the corresponding averages in the population.

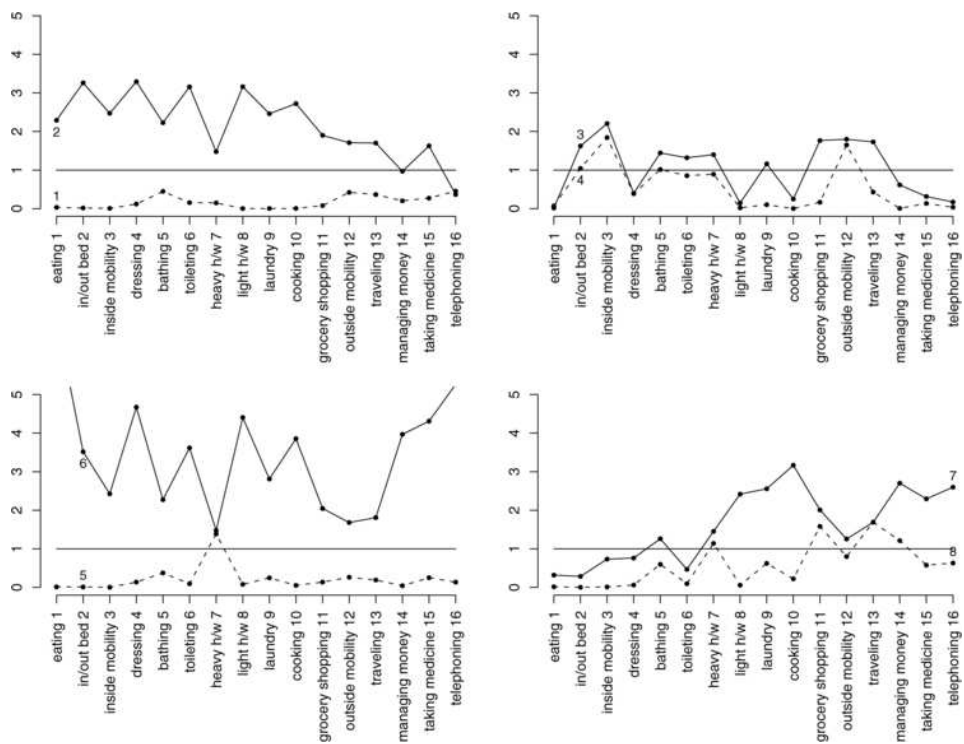

Fig. 3. *Functional disabilities relative frequencies for extreme profile pairs for the $K = 8$ GoM mixture model. Horizontal lines indicate average frequencies in the sample.*

Profiles $k = 3$ and $k = 4$ show patterns of frequencies that are somewhat similar to each other, indicating frequent difficulties with mobility activities, with profile 3 having noticeably higher frequencies on laundry, grocery shopping, and traveling.

Profile $k = 6$ is the profile of seriously disabled with most of the disability frequencies greater than 0.8 and greater than the corresponding average frequencies in the population. For a complete member of this profile, difficulties with each eating, dressing, light housework, managing money, taking medicine, and telephoning occur at least four times more often than in the NLTCS sample on average. Profile $k = 5$ points to low probabilities for most ADL and IADL items, but has a spike at the probability for doing heavy housework. An individual corresponding to this profile has difficulties with heavy housework one and a half times more often than the average chronically disabled person. This is a significant increase, given that the average frequency to experience difficulty with heavy housework is 0.68 in the NLTCS sample.

Profile $k = 8$ shows disability frequencies that are slightly higher than the average for heavy housework, grocery shopping, traveling and managing money. Profile $k = 7$ exhibits high frequencies for all IADL items, especially for those with significant cognitive components such as cooking, managing money and telephoning.

Having the profile interpretations at hand, we recall that they represent extreme types of chronically disabled individuals aged 65 and over. Apart from an estimated 15% of healthy individuals who have no disabilities with probability one, each (partially disabled) person in the population can be described through a vector of membership scores for the eight estimated profiles. Since the hyperparameter estimate $\hat{\alpha}_0 = 0.103$ is small, the posterior distribution of grades of membership is bathtub-shaped, which means that membership vectors are dominated by one component for a majority of individuals. Even though we focus on the population parameters in this paper, it is possible to use MCMC output to examine posterior distributions for each individual. One could also compute posterior estimates of various quantities of interest, such as the percentage of individuals in the sample that have membership vectors dominated by one profile (with $g_k > 0.95$, e.g.).

**9. Discussion.** Models that allow for specification of continuous latent constructs are increasingly popular among researchers in the social, behavioral, and health sciences since many latent variables of interest can be thought of as having fine gradations. When substantive theory justifies distinct latent categories as well as continuous latent variables, approaches that describe heterogeneity of individuals with respect to those discrete categories often focus on class membership probabilities. To give a few examples, Foody et al. (1992) emphasize the utility of posterior probabilities of class membership in the area of remote sensing; Muthen and Shedden (1999) model the class membership probability as a function of covariates in a study of alcohol dependence; Roeder, Lynch and Nagin (1999) address a similar issue by modeling uncertainty in latent class assignments in a criminology case study. The GoM model also addresses the issue of uncertainty in class membership, but by using a different approach that directly incorporates degrees of membership as model parameters.

Standard methods of estimating the GoM model described in Manton, Woodbury and Tolley (1994) do not rely on the GoM representation as a discrete mixture model and have questionable properties [Haberman (1995)], including instability of MLEs due to ridges in the likelihood function which are often present. The Bayesian GoM estimation algorithm developed originally in Erosheva (2002, 2003), on which the present paper is largely based, leans heavily on the structure provided by the latent class representation and has several advantages over likelihood-based estimation procedures for the GoM model. It is worth emphasizing one more time that the developed latent class representation of the GoM model places identical probability structure on observable variables and, hence, cannot possibly be distinguished from the continuous mixture GoM model on the basis of data [Erosheva (2006)].

Understanding the latent class representation of the GoM model, and thus viewing it as a special instance of individual-level or mixed membership models, makes it easier to establish direct connections with models from other areas. For example, although a clustering model with admixture developed for genetic data by Pritchard, Stephens and Donnelly (2000) and the standard GoM model appear to be quite different, they are both instances of the more general mixed-membership representation. The generalization is flexible enough to accommodate models for other data structures such as text documents [Erosheva, Fienberg and Lafferty (2004)].

Our goal for the NLTCS analysis in this paper was to explore the population characteristics of disability patterns as measured by the 16 ADL and IADL variables. Incorporation of covariates in the GoM modeling would be an obvious next step of great interest to social science researchers.

The preferred number of components identified by statistical criteria represents our best guess at the latent dimensionality in the NLTCS data under the GoM mixture model with a deterministically healthy compartment. Our choice of dimensionality is based on a number of assumptions, some of which may be worth exploring further. In particular, the assumption of local independence for the full set of ADL and IADL variables may be questionable. One possible approach to relax this assumption suggested by a reviewer is to focus on fitting the GoM models separately for the set of ADL and for the set of IADL variables, producing two sets of correlated GoM scores. Such a split-GoM model may turn out to be more appealing to disability researchers and to produce gains in interpretability. We expect to consider this and other forms of model simplification as we work toward our ultimate goal of developing a longitudinal version of the GoM model.

**Acknowledgments.** The authors would like to thank Matthew Stephens for insightful discussions, and Edo Airoldi and Tanzy Love for help in understanding the model selection problem.

E. A. EROSHEVA
DEPARTMENT OF STATISTICS
UNIVERSITY OF WASHINGTON
SEATTLE, WASHINGTON 98195-4322
USA
E-MAIL: elena@stat.washington.edu

S. E. FIENBERG
DEPARTMENT OF STATISTICS
CARNEGIE MELLON UNIVERSITY
PITTSBURGH, PENNSYLVANIA 15213-3890
USA
E-MAIL: fienberg@stat.cmu.edu

C. JOUTARD
GREMAQ - UNIVERSITÉ TOULOUSE 1
MANUFACTURE DES TABACS
21 ALLÉE DE BRIENNE
31000 TOULOUSE
FRANCE
E-MAIL: joutard@cict.fr